\documentclass[aps,prl,twocolumn,showpacs,amsmath,amssymb]{revtex4}
\usepackage{graphicx}
\usepackage{amsmath}
\usepackage{amsbsy}
\usepackage{dcolumn}
\usepackage{bm}
\usepackage{epsfig}

\begin{document}

\title{
Simultaneous ferromagnetic metal-semiconductor transition 
in electron-doped EuO}

\date{\today}
\author{Michael Arnold and Johann Kroha}
\affiliation{Physikalisches Institut, Universit\"at Bonn,
Nussallee 12, 53115 Bonn, Germany}

\begin{abstract}
We present a general framework to describe the simultaneous 
para-to-ferromagnetic and 
semiconductor-to-metal transition in electron-doped EuO.  
The theory correctly describes detailed experimental features of the
conductivity and of the magnetization, in particular the 
doping dependence of the Curie temperature. The existence of 
correlation-induced local moments on the impurity sites is 
essential for this description.
\end{abstract}
\bigskip
\noindent 
\maketitle


At room temperature stoichiometric europiumoxide (EuO) is a paramagnetic
semiconductor which undergoes a ferromagnetic (FM) transition at the Curie
temperature of $T_C=69~{\rm K}$. Upon electron doping, either by O defects
or by Gd impurities, this phase transition turns into a
simultaneous ferromagnetic and semiconductor-metal (SM) transition with nearly
100 \% of the itinerant charge carriers polarized and a sharp resistivity
drop of 8 to 13 orders of magnitude, 
depending on sample quality \cite{oliver1,oliver2,penney,steeneken}. 
Concomitant with this transition is a huge
colossal magnetoresistance (CMR) effect \cite{shapira}, 
much larger than in the intensely studied manganates \cite{tokura}.
These extreme properties make electron-doped EuO interesting for
spintronics applications. Known since the 1970s, these features have
therefore recently stimulated more systematic experimental 
studies with modern techniques and improved sample 
quality \cite{steeneken,ott,schmehl} as well as theoretical calculations 
\cite{schiller,sinjukow}. 

In pure EuO the FM ordering is driven by the Heisenberg 
exchange coupling between the localized Eu 4$f$ moments with
spin $S_f=7/2$ \cite{lee}. Upon electron doping, above $T_C$, 
the extra electrons are bound in defect
levels situated in the semiconducting gap, and the 
transition to a FM metal occurs when the majority states of the
spin-split conduction band shift downward to overlap with the
defect levels. 
Although this scenario is widely accepted, several questions of 
fundamental as well as applicational relevance have remained poorly
understood. (1) Why does the magnetic ordering of the Eu 4$f$ system
occur simultaneously \cite{steeneken} with the SM 
transition of the conduction electron system? 
(2) What is the order of the transition?
While the magnetic ordering of the 4$f$ system should clearly be of
2nd order, the metallic transition requires a {\it finite} shift of 
the conduction band and, hence, seems to favor a 1st order transition.
(3) How can the critical temperature $T_C$ be enhanced by doping for
spintronics applications? While in the Eu-rich compound 
EuO$_{1-x}$ a systematic $T_C$ increase due to the O defects (i.e.
missing O atoms) is not observed experimentally \cite{oliver1,oliver2}, 
a minute Gd doping concentration significantly
enhances $T_C$ \cite{matsumoto,ott}. 
An O defect in EuO$_{1-x}$ essentially binds the two excess electrons from 
the extra Eu 6s orbital and, therefore, should not carry a
magnetic moment. As shown theoretically in Ref.~\cite{sinjukow}, 
the presence of O defects with two-fold electron occupancy does not enhance 
$T_C$, in agreement with experiments \cite{oliver1,oliver2}.
In the present work we focus on the Gd-doped system Eu$_{1-y}$Gd$_y$ and 
calculate the temperature and 
doping dependent magnetization and resistivity from a microscopic model.
We find that the key feature for obtaining a $T_C$ enhancement is that
the impurities not only donate electrons but also carry a local magnetic 
moment in the paramagnetic phase.


{\it The model.} --- 
A Gd atom substituted for Eu does not alter the $S_f=7/2$ local 
moment in the Eu Heisenberg lattice but donates one dopant electron,
which in the insulating high-temperature phase is 
bound in the Gd 5d level located in the gap. 
Therefore, the Gd impurities are Anderson impurities with a local 
level $E_d$ below the chemical 
potential $\mu$ and a {\it strong} on-site Coulomb 
repulsion $U>\mu - E_d$ which restricts their electron occupation 
essentially to one. The hybridization $V$ with the conduction band is 
taken to be site-diagonal because of the localized Gd 5d orbitals. 
The Hamiltonian for the Eu$_{1-y}$Gd$_y$O system then reads,
\begin{eqnarray}
\label{hamiltonian}
H&=&\sum_{{\bf k}\sigma}\varepsilon_{{\bf k}} 
c_{{\bf k}\sigma}^{\dagger}c_{{\bf k}\sigma}^{\phantom{\dagger}}+H_{cd}+H_{cf}\\
\label{Hcd}
H_{cd}&=&E_{d} \sum_{i=1 \dots N_I,\sigma}
d_{i\sigma}^{\dagger}d_{i\sigma}^{\phantom{\dagger}} 
+  V \sum_{i=1 \dots N_I,\sigma}
(c_{i\sigma}^{\dagger} d_{i\sigma}^{\phantom{\dagger}} 
+ H.c.)\nonumber\\
&+& U \sum_{i=1 \dots N_I} d_{i\uparrow}^{\dagger} d_{i\uparrow}^{\phantom{\dagger}} 
    d_{i\downarrow}^{\dagger} d_{i\downarrow}^{\phantom{\dagger}} \\ 
\label{Hcf}
H_{cf}&=&-  \sum_{i,j} J_{ij} \vec S_{i}\cdot\vec S_{j} 
        - J_{cf}\sum_{i}\vec \sigma_{i}\cdot\vec S_{i} \ ,
\end{eqnarray}
where the first term in Eq.~(\ref{hamiltonian}) denotes conduction 
electrons with spin $\sigma$. 
The Eu 4$f$ moments $\vec S_i$ on the 
lattice sites $i=1,\dots, N$ are described in terms of a Heisenberg
model $H_{cf}$ with FM nearest and next-nearest neighbor couplings 
$J_{ij}$ and an exchange coupling $J_{cf}$ to the conduction electron 
spin operators at site $i$, $\vec\sigma_{i}=(1/2)\sum_{\sigma\sigma'}
c_{i\sigma}^{\dagger}\vec\tau_{\sigma\sigma'}c_{i\sigma'}^{\phantom{\dagger}}$, with $c_{i\sigma}=\sum_{\bf k} \exp(i{\bf k x_i})\,c_{{\bf k}\sigma}$ and 
$\vec \tau_{\sigma\sigma'}$ the vector of Pauli matrices. 
The Gd impurities at the random positions 
$i=1, ..., N_I$ are described by $H_{cd}$. For the numerical
evaluations we take $U\to\infty$ for simplicity.

For the present purpose of understanding the general form of the 
magnetization $m(T)$ and the systematic 
doping dependence of $T_C$ it is sufficient
to treat the 4$f$ Heisenberg lattice, $H_{cf}$, on mean field level,
although recent studies have shown that Coulomb correlations in the 
conduction band can soften the spin wave spectrum in similar systems 
\cite{golosov,perakis}. The effect of the latter on $m(T)$ can be absorbed in
the effective mean field coupling of the 4$f$ system,
$J_{4f} \equiv \sum_{j}J_{ij}$. We therefore choose $J_{4f}$ such that
for pure EuO it yields the experimental value of $T_C=69~{\rm K}$ 
\cite{oliver1,oliver2,shapira,steeneken}. 
For simplicity, we don't consider a direct coupling $J_{df}$ between the 
4$f$ and the impurity spins, since this would essentially
renormalize $J_{cf}$ only.  
The indirect RKKY coupling will also be neglected, 
since for the small conduction band fillings relevant here 
it is FM, like $J_{ij}$, but much smaller than $J_{ij}$.

In the evaluations we use a semi-elliptical bare conduction band density
of states (DOS) with a half width $D_0=8\, {\rm eV}$, 
(consistent with experiment \cite{steeneken}),
centered around $\Delta _0\approx 1.05\, D_0$ 
above the (bare) defect level $E_d$. 
The other parameters are taken as 
$J_{4f} \equiv \sum_{j}J_{ij} =  7\cdot 10^{-5} D_{0}$,
$J_{cf}=0.05 D_{0}$, $E_{d}=-0.4 D_{0}$,
and $\Gamma=\pi V^{2}=0.05 D_{0}^{2}$, 
where $J_{cf}\gg J_{4f}$ because $J_{4f}$ involves a non-local 
matrix element.


{\it Selfconsistent theory.} ---
The averaging over the random defect positions is done 
within the single-site $T$-matrix approximation, sufficient
for dilute impurities. This yields for the retarded conduction electron 
Green`s function $G_{c\sigma}({\bf k},\omega)$ in terms of its 
selfenergy $\Sigma _{c\sigma}(\omega)$,
\begin{eqnarray}
&&G_{c\sigma}({\bf k},\omega)=\left[\omega+\mu-\varepsilon_{\bf k}-\Sigma_{c\sigma}(\omega)\right]^{-1} \label{gc}\\
&&\Sigma_{c\sigma}(\omega)=n_{I} |V|^{2}G_{d\sigma}(\omega) -J_{cf}\langle S \rangle \sigma \label{se}
\end{eqnarray} 
where $G_{d\sigma}(\omega)$ is the defect electron propagator and 
$\langle S \rangle$ the average 4$f$--moment per site. 
In mean field theory it is obtained, together with the 
conduction electron magnetization $m$, as 
\begin{eqnarray}
&&\langle S \rangle = \frac{\sum_{S} S e^{-\beta(2J_{4f}\langle S \rangle + J_{cf}m)S}}{\sum_{S}e^{-\beta(2J_{4f}\langle S \rangle + J_{cf}m)S}}\\
&&m=\frac{1}{2}\int d \omega  f(\omega) [A_{c\uparrow}(\omega) - 
A_{c\downarrow}(\omega)]\label{magn}
\end{eqnarray}
where $f(\omega)$ is the Fermi distribution function and
$A_{c\sigma}(\omega)=- \sum_{{\bf k}} 
{\rm{Im}} G_{c\sigma}(k,\omega)/\pi$ the conduction electron DOS 
of the interacting system. 
\begin{figure}[t]
\scalebox{0.35}{\includegraphics[clip]{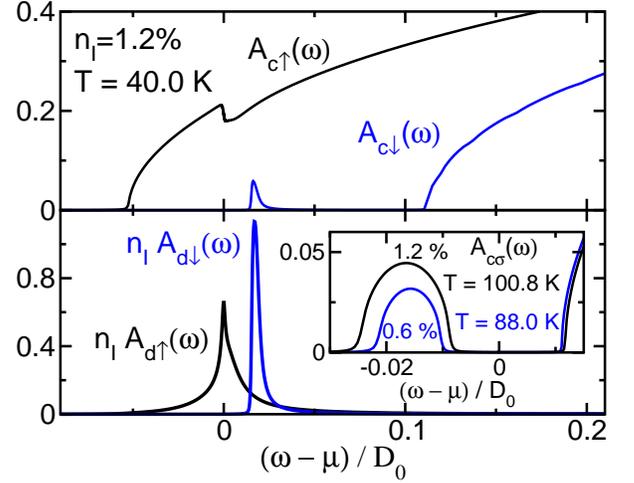}}\\
\caption{\label{fig1} (Color online) 
Conduction (upper panel) and impurity (lower panel) electron
DOS per lattice site for $T \ll T_C$. 
The impurity concentration is $n_{I}=1.2\%$ . The metallic
phase is fully spin polarized. Inset: $c$--electron DOS in the 
paramagnetic phase for $n_I=1.2\%$, $T=100.8\,{\rm K}$ and for
$n_I=0.6\%$, $T=88.0\,{\rm K}$. The chemical potential lies in the gap. 
}
\vspace*{-0.4cm}
\end{figure}
In order to treat the 
strongly correlated spin and charge dynamics of the Anderson impurities
without double occupancy beyond the static approximation, we  
use a slave particle 
representation and employ the non-crossing approximation 
(NCA) \cite{grewe}. For EuO the DOS at the Fermi level is so low or
even vanishing that the Kondo temperature is well below $T_C$ and 
Kondo physics plays no role. In this high-energy regime the NCA has been shown
to give quantitatively reliable results \cite{costi}. This remains
true even for a finite magnetization, where the NCA would develop 
spurious potential scattering singularities near $T_K$ only \cite{kirchner}.
One obtains the following set of equations for $G_{d\sigma}(\omega)$ 
in terms of the auxiliary fermion and boson propagators $G_{f\sigma}$, $G_{b}$,
their spectral functions $A_{f\sigma}$, $A_{b}$ 
and their selfenergies $\Sigma_{f\sigma}, \Sigma_{b}$,
\begin{eqnarray}
\Sigma_{f\sigma}(\omega)&=&\Gamma \int {d\varepsilon}\left[1-f(\varepsilon)\right] A_{c\sigma}(\varepsilon)G_{b}(\omega-\varepsilon )\label{sigmaf}\\
\Sigma_{b}(\omega)&=&\Gamma \sum_{\sigma}\int {d\varepsilon} f(\varepsilon) A_{c\sigma}(\varepsilon)G_{f\sigma}(\omega+\varepsilon )\label{sigmab}\\
\nonumber
G_{d\sigma}(\omega)&=&\int \frac{d\varepsilon} {e^{\beta \varepsilon}} \left[ G_{f\sigma}(\omega+\varepsilon )A_{b}(\varepsilon)-A_{f\sigma}(\varepsilon)G^{*}_{b}(\varepsilon-\omega)\right] \\
\label{Gd}
\end{eqnarray}
Note that in Eqs.~(\ref{sigmaf}, \ref{sigmab}) 
$A_{c\sigma}(\varepsilon)$ is the interacting DOS,
renormalized by the dilute concentration of Anderson impurities 
and the 4$f$--spins according to Eq.~(\ref{gc}).
For details of the NCA and its evaluation see \cite{costi}. 
The equations (\ref{gc}-\ref{Gd})  
form a closed set of selfconsistent integral equations.
They are solved iteratively, fixing the total electron 
number per lattice site in the system, 
\begin{eqnarray}
n= \sum_{\sigma}\int \!d\omega  f(\omega)\, 
\left[A_{c\sigma}(\omega)+n_I\,A_{d\sigma}(\omega)\right]=n_I
\label{pnumber}
\end{eqnarray}
by the chemical potential $\mu$ in each step.

\begin{figure}[t]
\scalebox{0.45}{\includegraphics[clip]{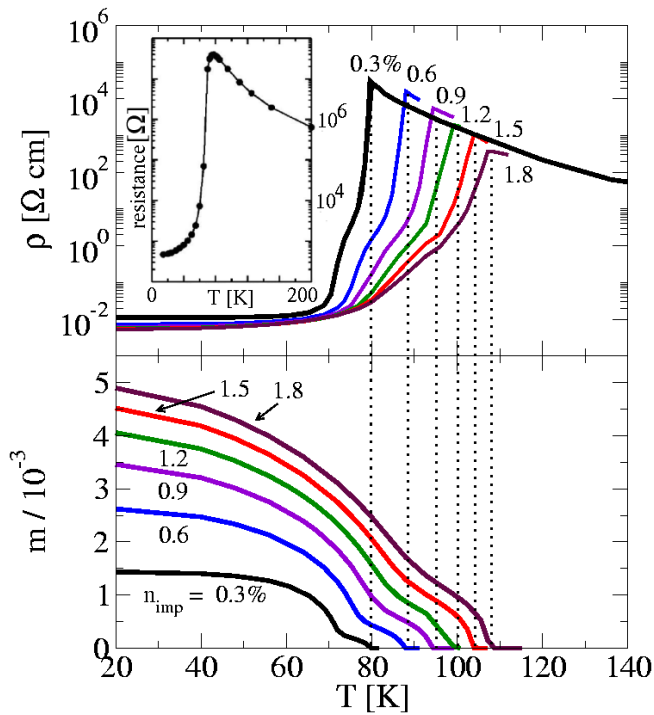}}\\
\caption{\label{fig2} (Color online)
Simultaneous FM and semiconductor-metal transition as seen in the 
$T$--dependence of the magnetization $m$ and of the resistivity
$\rho=1/\sigma$. The Curie temperature depends significantly on the 
impurity concentration $n_{I}$. The inset shows a typical 
experimental $\rho(T)$ curve, taken from Ref.~\cite{steeneken}.}
\vspace*{-0.4cm}
\end{figure}


{\it Electrical conductivity.} ---
The current operator $\hat{\bf j}$ can be derived from the continuity
equation, $\partial\hat\rho_i/\partial t + \nabla\ \cdot \hat{\bf j} =0$,
and the Heisenberg equation of motion for the total local charge operator
$\hat\rho_i$ at site $i$. Because the impurity Hamiltonians $H_{cf}$, $H_{df}$
conserve $\hat\rho_i$, only $c$--electrons contribute to the current, and
one obtains \cite{schweitzer},
$
\hat{\bf j}=({e}/{\hbar}) \sum_{{\bf k}\sigma}{\partial \varepsilon_{\bf k}}/
  {\partial {\bf k}} \
c_{{\bf k}\sigma}^{\dagger} c_{{\bf k}\sigma}^{\phantom{\dagger}}
$.
The linear response conductivity then reads for a local 
selfenergy \cite{schweitzer},
\begin{equation}
\sigma=\frac{\pi e^{2}}{3 \hbar V} \sum_{{\bf k}\sigma} \int d\omega \left(
  -\frac{\partial f}{\partial \omega} \right)  A_{c\sigma}^{2}({\bf k},\omega)
\left( \frac{\partial \varepsilon_{\bf k}}{\partial {\bf k}} \right)^{2} \ .
\label{cond1}
\end{equation}

\begin{figure}[t]
\scalebox{0.31}{\includegraphics[clip]{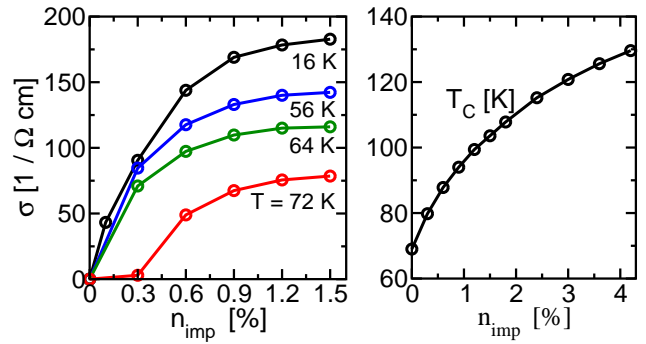}}\\
\caption{\label{fig3} Conductivity $\sigma$
for various temperatures (left panel) and Curie temperature $T_C$
(right panel) as function of impurity concentration $n_I$. 
The data points at $n_{I}=0$ in the left panel are extrapolations.}
\vspace*{-0.4cm}
\end{figure}

{\it Results and discussion.} --- 
The results of the selfconsistent theory, Eqs.~(\ref{gc}--\ref{pnumber}),
and for the conductivity, Eq.~(\ref{cond1}), are presented in 
Figs.~\ref{fig1}--\ref{fig3}. They allow to draw a complete picture of the
FM semiconductor-metal transition in Gd-doped EuO. The spectral
densities per lattice site
above and below the transition are shown in Fig.~\ref{fig1}.
In the paramagnetic, insulating phase the hybridization between 
$d$-- and $c$--electrons necessarily implies the appearance of a 
conduction electron sideband (Fig.~\ref{fig1}, inset), situated 
below $\mu$ and at the same energies inside the semiconducting gap 
as the impurity $d$--band. The $d$-band (not shown) has a similar 
width and shape as the $c$-sideband. The combined weight of the 
$c$--sideband and the $d$-band adjusts itself selfconsistently 
such that it just accommodates the total electron number,
$n=n_I$. Note that the weight of the $d$--band per impurity and spin 
is $\lesssim 1/2$, because the doubly occupied weight is shifted to
$U\to \infty$ \cite{costi}.  

The $c$--4$f$ exchange coupling $J_{cf}$  induces an effective 
FM coupling between the electrons of the $c$--$d$ system. Hence,
either the 4$f$-- or the $c$--$d$--electron system can drive a
FM transition, depending on which of the (coupled) subsystems has the 
higher $T_C$. We have chosen $J_{cf}$ (see above) large enough that 
the transition is driven by the $c$--$d$--electrons, because this 
will yield detailed agreement with the experiments 
\cite{steeneken,ott,matsumoto}. 
In this case, $T_C$ is naturally 
expected to increase with the impurity density $n_I$. The results
for the $T$-dependent conduction electron magnetization $m(T)$,
Eq.~(\ref{magn}), and for the doping dependence of $T_C$ are 
shown in Fig.~\ref{fig2}, lower panel, and in 
Fig.~\ref{fig3}, right panel, respectively. It is seen that not only
$T_C$ increases with the impurity concentration, in agreement
with recent measurements on Eu$_{1-y}$Gd$_{y}$O$_{1-x}$ \cite{matsumoto,ott},  
but also that $m(T)$ has a dome-like tail near $T_C$, before it increases
to large values deep inside the FM phase. 
From our theory this feature is traced back 
to the mean-field-like 2nd order FM transition of the electron system,
while the large dome in the magnetization further below $T_C$ is 
induced by the FM ordering of the 4$f$ system, whose magnetization 
is controlled by $J_{4f}$ and sets in at lower $T$. This distinct feature is 
again in agreement with the experimental findings \cite{matsumoto,ott}
and lends significant support for the present model for Eu$_{1-y}$Gd$_{y}$O. 
We note that the Eu-rich EuO$_{1-x}$ samples  of 
Ref.~\cite{matsumoto} also show a magnetization tail and a $T_C$ enhancement,
suggesting (small) magnetic moments on the O defects. However, the
nature of the O defects requires further experimental and theoretical studies.  
The conduction electron polarization $P(T)=m(T)/n_c(T)$
does not show this double-dome structure and below $T_C$ 
increases steeply to $P=1$ (not shown in Fig.~\ref{fig2}). 
The FM phase is connected with a spin splitting of the 
$c$-- as well as the $d$--densities of states, as shown in Fig.~\ref{fig1}.
The narrow $d$-band induces a Fano dip structure in the $c$
majority band and a small sideband in the $c$ minority band.
Note that for the present scenario the existence of preformed local
moments on the impurities, 
induced by strong Coulomb repulsion $U$, is essential. 
Without these moments the transition of the electron system
would be purely Stoner-like, and, because of the extremely low 
conduction electron DOS at the Fermi level, its $T_C$ would be far
below the Curie temperature of the 4$f$ system, so that no doping 
dependence would be expected \cite{sinjukow}.  

\begin{figure}[t]
\scalebox{0.31}{\includegraphics[clip]{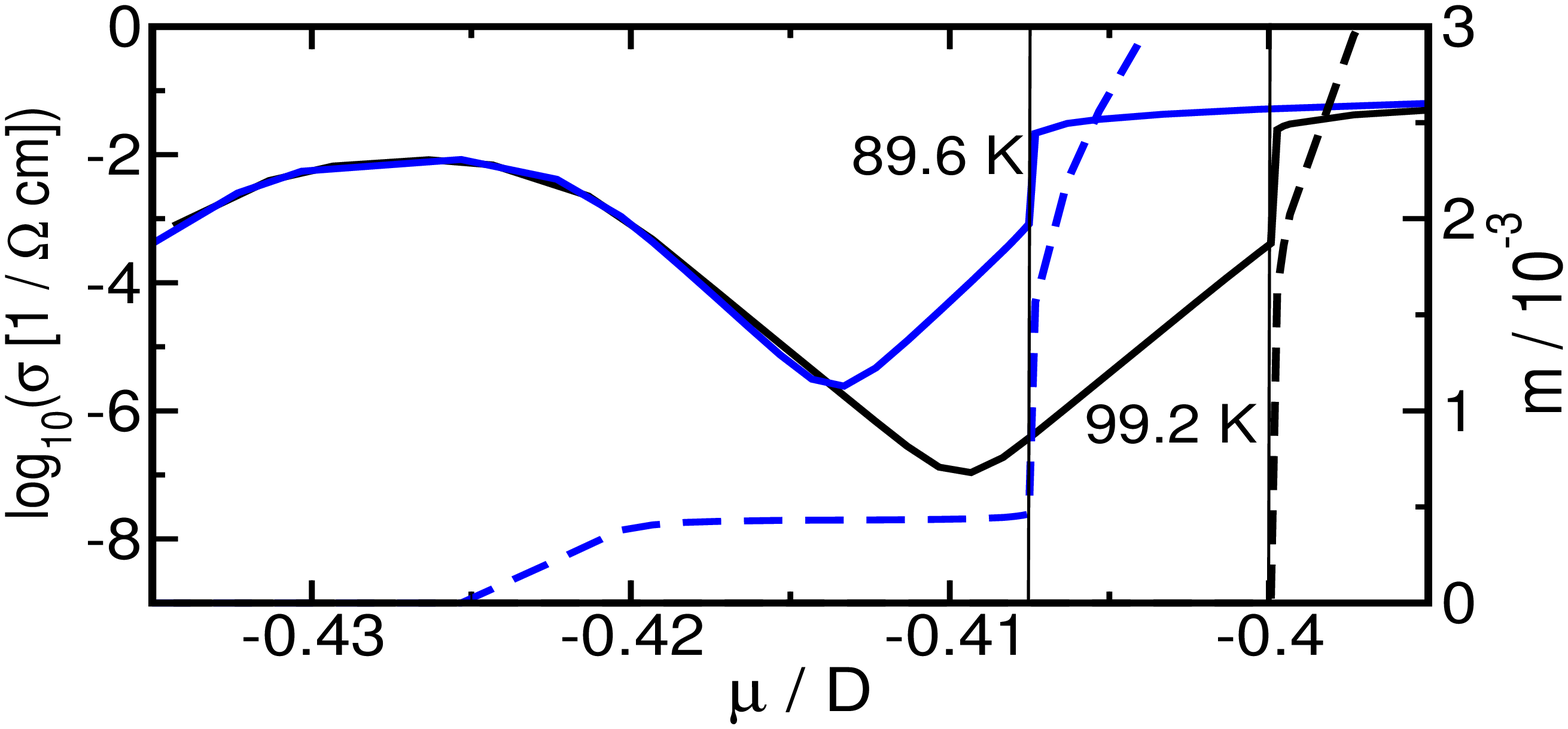}}\\
\caption{\label{fig4} (Color online) 
Conductivity (---) and magnetization 
(- - -) as a function of $\mu$ for 
$T=89.6\,{\rm K}$ and $T=99.2\,{\rm K}$. 
At the ungated points, $\mu \approx -0.414\, D_{0}$ for $T=89.6\,{\rm K}$ 
and $\mu \approx -0.410\, D_{0}$ for $T=99.2\,{\rm K}$, the 
total electron number per lattice site is equal to the impurity concentration,
$n=n_I$.}
\vspace*{-0.2cm}
\end{figure}

We now discuss the conductivity and the 
simultaneity of the FM and the SM transitions. 
In the paramagnetic phase, the system is weakly semiconducting, because
$\mu$ lies in the gap (Fig.~\ref{fig1}, inset). 
When the FM transition occurs, the impurity d-band must acquire a spin 
splitting in such a way that at least part of the minority $d$--spectral
weight lies above the chemical potential 
$\mu$, in order to provide a finite magnetization.
Since near the transition the spin splitting is small, the majority
$d$--band must, therefore, also be shifted to have 
overlap with $\mu$ (Fig.~\ref{fig1}),
and so must the hybridization-induced $c$-electron sideband 
(which eventually merges with the main conduction band for $T$ sufficiently
below $T_C$). This immediately implies a transition to a metallic state,
simultaneous with the FM transition, as seen in Fig.~\ref{fig2}.
Because of the small, but finite thermal occupation of the states
around $\mu$, we find that this shifting of spectral weight
occurs continuously, which implies the FM semiconductor-metal
transition to be of 2nd order (see Fig.~\ref{fig2}). 
The doping $n_I$ dependence of the conductivity is shown in Fig,~\ref{fig3},
left panel. It is seen that the metallic transition can be driven 
by increasing $n_I$, if $T>T_C(n_I=0)$.

As an alternative to Gd-doping the charge carrier concentration
$n$ can be controlled independently of the impurity concentration $n_I$
by varying the chemical potential $\mu$, e.g. by
applying a gate voltage to an EuO thin film. The conductivity $\sigma$ and
magnetization $m$ as a function of $\mu$ are shown in Fig.~\ref{fig4} for two
temperatures. To both sides of the ungated system ($n=n_I$)  
$\sigma$ increases exponentially upon changing $\mu$, characteristic
for semiconducting behavior. By increasing $\mu$, the FM-metallic 
transition is finally reached. I.e. the magnetization can be switched,
in principle, by a gate voltage. The non-monotonic behavior of $\sigma$
towards more negative $\mu$ reflects the energy dependence of the 
$c$ sideband. A more detailed study will be presented elsewhere.

To conclude, our theory indicates that 
in Gd-doped EuO the existence of preformed local moments on
the impurity levels inside the semicondicting gap is essential
for understanding the distinct shape of the magnetization $m(T)$ 
near the ferromagnetic semiconductor-metal transition. 
The FM ordering is driven by these impurity moments which are 
superexchange coupled via the 4$f$ moments of the underlying 
Eu lattice. This scenario immediately implies an increase of the
Curie temperature with the impurity concentration, in agreement
with experiments. The double-dome shape of $m(T)$ arises because of
the successive ordering of the dilute impurity and of the dense
Eu 4$f$ systems, as $T$ is lowered. 
The dynamical 
accumulation of conduction spectral weight at the chemical potential,
induced by the hybridization $V$ and the constraint of an emerging
magnetization at the FM transition, 
implies the FM and the SM transition to be simultaneous and of 2nd order. 
The magnetization can be switched by
applying a gate voltage. This might be relevant for spintronics
applications.

We wish to thank T. Haupricht, H. Ott, and H. Tjeng for useful discussions.
J.K. is grateful to the Aspen Center for Physics
where this work was completed. This work is supported by DFG through SFB 608.

\end{document}